\documentclass[12pt]{article}
\usepackage{amssymb,epsf}
\def\lsim{\mathrel{\rlap {\raise.5ex\hbox{$ < $}}
{\lower.5ex\hbox{$\sim$}}}}
\def\gsim{\mathrel{\rlap {\raise.5ex\hbox{$ > $}}
{\lower.5ex\hbox{$\sim$}}}}
\def\sqr#1#2{{\vcenter{\vbox{\hrule height.#2pt
        \hbox{\vrule width.#2pt height#1pt \kern#1pt
           \vrule width.#2pt}
        \hrule height.#2pt}}}}

\def\@vacuum{}
\def\draftmarginnote#1{\marginpar{\raggedright\scriptsize\tt#1}}
\def\draft{\oddsidemargin -.2truein
        \def\@oddfoot{\sl preliminary draft \hfil
        \rm\thepage\hfil\sl\today\quad\militarytime}
        \let\@evenfoot\@oddfoot \overfullrule 3pt
        \let\label=\draftlabel
        \let\marginnote=\draftmarginnote
   \def\@eqnnum{(\theequation)\rlap{\kern\marginparsep\tt\@eqnlabel}%
\global\let\@eqnlabel\@vacuum}  }
\def\subequations{\refstepcounter{equation}%
  \edef\@savedequation{\the\c@equation}%
  \@stequation=\expandafter{\theequation}
  \edef\@savedtheequation{\the\@stequation}
  \edef\oldtheequation{\theequation}%
  \setcounter{equation}{0}%
  \def\theequation{\oldtheequation\alph{equation}}}

\def\endsubequations{\setcounter{equation}{\@savedequation}%
  \@stequation=\expandafter{\@savedtheequation}%
  \edef\theequation{\the\@stequation}\global\@ignoretrue
  \vspace*{-12pt} \\}
\def\bs{\begin{subequations}}
\def\es{\end{subequations}}
\relax

\def\thefootnote{\fnsymbol{footnote}}
\def\be{\begin{equation}}
\def\ee{\end{equation}}
\def\ba{\begin{eqnarray}}
\def\ea{\end{eqnarray}}

\def\im{\, {\rm Im}\, \tau}
\def\iT{{\,{\rm Im}\, }T}
\def\iU{{\,{\rm Im}\, }U}

\def\ee{\end{equation}}
\def\bea{\begin{eqnarray}}
\def\eea{\end{eqnarray}}

\def\np#1#2#3{Nucl. Phys. {\bf{B#1}} (#2) #3} 
\def\pl#1#2#3{Phys. Lett. {\bf{B#1}} (#2) #3}

\def\pr#1#2#3{Phys. Rev. {\bf{D#1}} (#2) #3}

\newcommand{\uarrw}[0]{\mathrel{
{\raise.5ex\vbox{\hrule width 1cm}\hskip-6pt\rightarrow}}}
\def\thebibliography#1{%
\vskip 0.5cm \centerline{\bf References}
\list{%
[\arabic{enumi}]}{\settowidth\labelwidth{[#1]}
\leftmargin\labelwidth
\advance\leftmargin\labelsep
\usecounter{enumi}}
\def\newblock{\hskip .11em plus .33em minus .07em}
\sloppy\clubpenalty4000\widowpenalty4000
\sfcode`\.=1000\relax}

\renewcommand{\theequation}{\arabic{section}.\arabic{equation}}

\begin{document}
\renewcommand{\theequation}{\arabic{section}.\arabic{equation}}
\begin{titlepage}
\begin{flushright}LPTENS-00/23,\\ 

hep-th/0005271 \\
\end{flushright}
\begin{centering}
\vspace{1.0in}
\boldmath
{\bf \large Conformal  String Sector of  M-Theory }
\\
\unboldmath
\vspace{1.7 cm}
{\bf \large  Costas Kounnas} \\
\medskip
\vspace{.2in}
{ \it Laboratoire de Physique Th\'eorique, ENS, F-75231 Paris France }\\
                                     
\end{centering} 
\vspace{2.5cm}
{\bf Abstract}\\
We consider a conformal system of a  string and a particle 
defined in $D=10+2$ space-time dimensions. The extra time-like 
dimension is 
a gauge artifact and can be eliminated by choosing a gauge in which
the $SO(10,1)$ Lorentz symmetry is manifest. The effective theory
of string observables is the 11d supergravity. The same theory  
compactified on $T^2$ provides a non-perturbative unified picture
of the Type IIA, Type IIB and 11d supergravity. This is confirmed 
by explicit determination of the  $R^4$-terms  which are finite and 
manifestly
$SL(2,Z)$ invariant as expected by the  U-duality conjecture in 
nine non-compact dimensions with maximal supersymmetry.\\

\vspace{1cm}
\begin{centering}
{\it Talk given at the European Network meeting:\\
``Non-perturbative Effects and Symmetry in Quantum Field Theory''}\\
{$~~~~~~~~~~~~~~~~~~~~~~~~~~~~~~~~~~~~~~~~$
\it Paris 1-7 September 1999.}
\end{centering}
\vspace{1.5cm}
\hrule width 6.7cm
.\\
Research partially supported by the EEC under the contract
TMR-ERBFMRX-CT96-0045.\\
e-mail: kounnas@physique.ens.fr
\end{titlepage}

\newpage
\renewcommand{\thefootnote}{\arabic{footnote}}

\boldmath
\section{ Introduction}
\unboldmath

In space-time  dimension lower than ten, all supergravity
theories with maximal supersymmetry  have a  universal
massless sector, e.g. the gravity supermultiplet \cite{cj}.
Indeed, all supergravities with $N_{max}$ are identical modulo field 
redefinitions, which correspond to the vev's of scalars in the 
supergravity multiplet \cite{cj}, and after performing ``electric-magnetic" 
duality-like transformations acting on the gauge fields (classical 
U-duality transformations). 
These effective  theories can be  constructed either by compactification 
on a $T^{(n+1)}$ torus from the $N=1, d=11$ supergravity \cite{cj} or by 
string compactification on $T^{(n)}$ of the type IIA or type IIB 
ten dimensional superstrings.
This universality of the maximal  supergravities 
($N_{max}=8$ in four dimensions) leads to the conjecture that they 
 are all 
identical  at the non-perturbative level.  This suggests  the existence 
of a  more fundamental theory, (M-theory?) \cite{W1}, which is defined
necessarily in dimension higher than ten. 
In  particular, the eleven- dimensional  supergravity is the 
effective ``low-energy" local field theory of the would be fundamental
theory \cite{W1}.
Furthermore, the  universality conjecture for $N_{max}$ seems  to be
valid even for less supersymmetric theories  with 1/2 and 1/4
of  ${N_{max}}$ \cite{W1}--\cite{Fth}. 
This suggests further  
that  all superstring
theories  with the same number of supersymmetries in a given 
dimension are equivalent at the non-perturbative level. Thus,
 Heterotic $\leftrightarrow$
type I $\leftrightarrow$ type IIA
$\leftrightarrow$ type IIB,
must be connected in lower dimensions by perturbative and/or 
non-perturbative $U$-duality transformations \cite{W1}--\cite{Fth}. 
 
What can be the origin of the $U$-duality connections?
Here we will focus on a possible  geometrical origin
due to the presence of some extra ``hidden" dimensions which  
enable us to describe the  complete spectrum of all topological 
non-perturbative BPS states of the maximal  supersymmetric 
theories in nine dimensions.
In the past few  years, it has been realized that  one
``hidden" dimension is not enough to describe the spectrum of the
topological BPS states. This implies the existence of more
than one hidden dimension. The type IIB theory, for instance,
suggests a  fundamental theory (F-theory) in 10+2 dimensions \cite{Fth}. 
Vafa and others  suggested a working algorithm which extends in a
consistent way the type IIB non-perturbative BPS spectrum and the 
$U$-duality properties in lower dimensions.
If  there are more than one dimension and especially a time-like
one  how do we interpret them?
One of the possibilities is to define a  combined
system of a String and a Particle living in $D_{crit}=(10+2)$ 
dimensions \cite{BK1},\cite{BK2}.

We will shortly review this possibility in the next section
and then we will study in more details the $R^4$ terms of this
theory in a special  gauge where $SO(10,1)$ Lorentz invariance
is manifest  with one physical time. 

\boldmath
\section{ String \& Particle}
\unboldmath  

Consider a string and a particle \cite{BK1},\cite{BK2}
 described by a  world-sheet 
$X^\mu \left( \tau ,\sigma \right) $ and a world-line 
$Y^\mu \left( \tau \right) $
$$
S_{str}(X^{\mu}, A_{\alpha},g_{\alpha \beta}, Q_{\mu}) + 
S_{par}(Y, B, e, P_{\mu})+ Q\cdot P\,,
$$
\begin{eqnarray}
\end{eqnarray}

$$
S_{str}=\frac 12\int_0^Td\tau \int d\sigma \,\,\sqrt{-g}g^{\alpha \beta}
\left( \partial_{\alpha} X^\mu 
-P^\mu A_{\alpha}\right)
$$
\begin{eqnarray}
\times \left( \partial _\beta X^\nu -P^\nu
A_{\beta}\right) \eta _{\mu \nu },
\end{eqnarray}
$$
S_{part} =\frac 12\int_0^Td\tau \,\left[ e^{-1}\left( \partial _\tau Y^\mu
-Q^\mu B\right) ^2-e m ^2\right].
$$
\begin{eqnarray}
\end{eqnarray}
The two actions $S_{str}$ and $S_{part}$ are
invariant under independent reparametrizations on the world-sheet and
on the  world-line.
Then one can choose the usual conformal gauge for the string,
$\sqrt{-g}g^{\alpha \beta}$
=$\eta ^{\alpha \beta},$
and  $e=1$ for the particle.
The equations of motion
for $X^\mu \left( \tau ,\sigma \right), 
Y^\mu \left( \tau \right) ,$ in
the gauges we have chosen, are:
\begin{eqnarray}
\partial _{+}\left( D_{-}X^\mu \right) +\partial _{-}\left( D_{+}X^\mu
\right) =0,~ \partial _\tau q^\mu =0,~~
\end{eqnarray}
where
\begin{eqnarray}
D_{\pm }X^\mu  &=&\left( \partial _{\pm }X^\mu -P^\mu
A_{\pm}\right) ,\quad \nonumber \\
q^\mu  &=&\left( \partial _\tau Y^\mu -Q^\mu B\right).~ 
\end{eqnarray}
There are additional gauge invariances, which
may be understood in the spirit of  gauged WZW models
\begin{eqnarray}
\delta _1 X^\mu  &=&P^\mu \Lambda_1 \left( \tau ,\sigma \right)
,\quad \delta _1A_{\alpha}=\partial _{\alpha}\Lambda_1 
\left( \tau ,\sigma \right) , \nonumber\\
\delta _2Y^\mu  &=&Q^\mu \Lambda _2\left( \tau \right) ,\quad
\,\quad \delta _2B=\partial _\tau \Lambda _2\left( \tau \right).
\end{eqnarray}

We can  solve the constraints \cite{BK1},\cite{BK2} and the 
equations of motion (for $m\ne 0$)
\begin{eqnarray}
Q^\mu \sim \,p^\mu ,\quad P^\mu \sim q^\mu ,\quad
p\cdot q=0.
\end{eqnarray}
In the light-like case  $P^2=0=q^2$ ( $m=0$), the action 
$S_{str}\,$ has no $A_{+}A_{-}$ term, and acquires an
additional gauge symmetry:
\begin{eqnarray}
\delta _3~X^\mu \left( \tau ,\sigma \right) =0,
\quad \delta _3~A_{\pm }=\pm
\partial _{\pm }\Lambda _3\left( \tau ,\sigma \right).~~ 
\end{eqnarray}
In the background of the  {\it massless particle}, 
two string components, rather than only one, are eliminated 
by the gauge invariances and thus,
$\partial _{\pm }X^\mu-c_1q^\mu $,
has no components along the light-like
$q^\mu$ and $~q\cdot \partial _{\pm }X$
$=0$.
For a  {\it massive particle}, these two
conditions correspond to one and the same component
\cite{BK1},\cite{BK2}.

The equation of motion for the string is easily solved, 
since it has the free string form
$\partial _{+}\partial _{-}X^\mu =0.$
The general solution is given in terms of left- and right- movers
$$
X_{\mu } =X_{\mu }^{\left( +\right) }\left( \sigma ^{+}\right) +
X_{\mu }^{\left( -\right) }\left( \sigma ^{-}\right)+c_1(\sigma ^{+}
+\sigma ^{-}) q_{\mu},~
$$
$$
X_{\mu }^{\left( \pm \right) }\left( \sigma ^{\pm}\right) =
\frac 12\left( x{\mu }+
\frac{\sigma ^{\pm }}{2\pi }p_{\mu} \right)-i\sum_{n\neq 0}
\frac 1n\alpha _{n\mu }^{\left( \pm \right) }\,e^{in\sigma ^{\pm }}.~
$$
\begin{eqnarray}
\end{eqnarray}
The solution of the particle equation, 
\begin{eqnarray}
Y^\mu \left( \tau \right) =y^{\mu}+\left( q^\mu +c_2p^\mu \right) \tau,
\end{eqnarray}
shows that {\it the particle moves freely, except for the
orthogonality constraint} $p \cdot q=0$.
Due to this  constraint, at the quantum level there are 
anomalies in arbitrary space-time dimensions \cite{BK1}, \cite{BK2}; 
the String \& Particle 
System is anomaly free only in special dimensions depending on the
parameter. Furthermore,  $m=0$  when  super-repara\-metrization invariance
is present 
on the world-sheet of the string and on the world-line of the particle,
$m=0$.

$\bullet$ When $ m= 0$ $D_{cr}=28$, for the  bosonic  string
with $SO(26,2)$ Lorenzt  invariance.

$\bullet$ For superstring   $m= 0$, $D_{cr}=12$, with  $SO(10,2)$ 
Lorentz symmetry.\\\\
All quantization approaches, BRST quantization, 
Light-cone quantization or  gauge WZW approach,  give the
same critical dimensions \cite{BK1},\cite{BK2}.

\boldmath
\section{String \& Particle Partition function and $R^4$-terms}
\unboldmath
  
 There is a gauge choice which  solves the orthogonality
condition $q_{\mu}\partial_{\pm}X^{\mu}=0$ 
with $SO(1,10)$ covariance for all  string 
observables in the particle background with momentum $q_{\mu}$:
$$
q_{\mu}=(~\epsilon, 0 ~|~ q_{_{10}},q_{_9},~0,0,0,0,0,0,0,0~)
$$
$$
\partial_{\pm}X^{\mu}=(0, V^0~|~V_{\pm}^{10}, V_{\pm}^9,  ~V^{I}~ )
,~~I=1,...,8.
$$
\begin{eqnarray}
\end{eqnarray}

In this gauge, one of the  two time-like string coordinates is eliminated
and one is left with the physical time coordinate as in usual string
theories. Notice that two orthogonal time-like vectors {\it  cannot exist
in a space with a single time-like dimension}.
The initial  $SO(2,10)$ covariant constraints 
$q_{\mu}\partial_{\pm}X^{\mu}=0$ 
become in this gauge:
\begin{eqnarray}
q_i\partial_{\pm}X^{i}=0~,~~~~i=9,10.
\end{eqnarray}
These constraints eliminate the quantum excitations of the string
oscillators parallel to the vector $q_i$ like in the gauge  WZW-models. 
There is  however a remnant of the zero-mode topological sector
defined by the $9th$ and $10th$ left- and right- moving
compactified coordinates of the string.
The reduced (2,2) Lorentzian lattice
$\Gamma (2,2)_{q_i}(T,U)$ depends
on both the moduli $T$ and $U$:
$~~~T\sim iR_9 R_{10}$, $~~~~ U\sim iR_9/R_{10}.$
\begin{eqnarray}
\Gamma (2,2)_{q_{i}}= 
e^{-\pi\im  (\,p_iG^{ij}p_j+ n^iG_{ij}n^j\,)+
2i\pi{\rm Re}\tau\, p_in^i},
\end{eqnarray}
\begin{eqnarray}
p_i=m_i+B_{ij}n^j~{\rm with}~ p_iq^i=n^iG_{ij}q^j=0.~
\end{eqnarray}
The orthogonality constraint  projects the $\Gamma(2,2)$  lattice
on-to a sum of $\Gamma(1,1)$ sectors  according to the co-prime integers
$(p,q)$ defined as $q^i={\hat M}{\hat q}^i={\hat M}(p,q)$.
Then,
$$
\Gamma(2,2)_{q^i}= 
e^{-\pi\im  \left(\, M^2 {\hat q}_iG^{ij}{\hat q}_j 
+{N^2\over {\hat q}_i G^{ij} {\hat q}_j}\,\right)+
2i\pi{\rm Re}\tau MN} .
$$
\begin{eqnarray}
\end{eqnarray}
In terms of $T$ and $U$ moduli:
\begin{eqnarray}
{\hat q}_iG^{ij}{\hat q}_j= {1\over \iT}~{|p+qU|^2\over \iU}.
\end{eqnarray}
The sum over $(p,q)$-sectors yields to 
$SL(2,Z)_{U}$ invariant results. 

In the zero winding sector $N=0$ the sum over $M$ and $(p,q)$ 
can be reorganized to {\it two} Kaluza-Klein momenta,
$$ 
m_i=M \epsilon_{ij}{\hat q^j} 
$$
\begin{eqnarray}
\Gamma({2,2})_{q^i}|_{(N=0)}=
e^{-\pi\im (\, m_iG^{ij} m_j\,)}.
\end{eqnarray}
After Poisson re-summation on $m_i$:
\begin{eqnarray}
\Gamma({2,2})_{q^i}|_{(N=0)}={ \iT \over \im }
e^{-\pi  (\, {\tilde m}^iG_{ij} 
{\tilde m}^j\,) \over \im}.
\end{eqnarray}
This result has an obvious eleven dimensional interpretation.
Thus the sum over all $(p,q)$ string sectors  give rise to two
instead of one Kaluza-Klein momenta, with an obvious eleven-dimensional 
interpretation.

We can go even further and calculate the $R^4$-gravitational corrections.
This can be easily done using the techniques developed in 
refs\cite{kk},\cite{gkpr}
where the flat space-time is replaced 
by a non trivial gravitational background  deformed
by the left- and right- helicity operators \cite{kk},\cite{gkpr}
$~vQ_{left},~ \bar{v}Q_{right}~$ with:

$$ R \ne 0\rightarrow v,\bar{v}\ne 0. $$

From the deformed  partition function \cite{kk},\cite{gkpr}:
\begin{eqnarray}
Z(\tau,\bar{\tau} |v,\bar{v}) =
{(\im)^{-1} \over  \eta^2(v)\,\bar{\eta}^2(\bar{v})}~
{(\im)^{-5/2}\over \eta^{5}~ \bar{\eta}^{5}}~
{\Gamma_{2,2}(p,q) \over \eta ~\bar{\eta}}\nonumber
\\
\times~ {1 \over 2}~ \sum_{a,b} (-)^{a+b+ab}{\vartheta[^a_b](v)
~\vartheta[^a_b]^3 \over  \eta^4}\nonumber 
\\
\times~ {1 \over 2} ~\sum_{\bar{a},\bar{b}}
(-)^{\bar{a}+\bar{b}+\bar{a}\bar{b}}
~{ \bar{\vartheta}[^{\bar a}_{\bar b}](\bar v)
~\bar{\vartheta}[^{\bar a}_{\bar b}]^3
\over \bar{\eta}^4}.\nonumber\\
\end{eqnarray}
one obtains the  $R^4$-term from the term which is proportional
to  $v^4\,\bar{v}^4$ -term. Details of these technics appear in
refs\cite{kk},\cite{gkpr}. Here we display  and give some comments 
on our  results. 
\begin{eqnarray}
F_{R^4}=\int_{\cal F}~{\partial \tau \partial \bar{\tau} \over t^2 }~ 
t^{1/2}~\Gamma_{2,2}(p,q),
\end{eqnarray}

$$
F_{R^4}=\iT \left( \int_{\cal F}~
{ \partial\tau \partial\bar{\tau} \over t^2 }
+\int_0^{\infty}{ dt \over t^{5/2}}~ 
e^{-\pi { {\tilde m}^i G_{ij} {\tilde m }^j  \over t }} \right).
$$
\begin{eqnarray}
\end{eqnarray}

In the second term ${\tilde m}^i \ne 0$
($t=\im $).

$$
F_{R^4}=\iT \left( {\pi \over 3} +{\iT^{-3/2} \over 2\pi}
\sum_{{\tilde m}^i} 
{\iU^{3/2} \over |{\tilde m}^1 + {\tilde m}^2 U|^3 } \right)
$$
\begin{eqnarray}
= \iT~{\pi \over 3} + \iT^{-1/2}
{\zeta(3)\over 2\pi}~E_{3/2}(U),~~ 
\end{eqnarray}
where $E_{3/2}$ is the Eisenstein series with weight $w$=$3/2$:

$$
E_{3/2}(U)=\sum_{p,q} {\iU^{3/2}\over |p+qU|^3},~~~(p,q)~
{\rm coprimes}.~
$$
\begin{eqnarray}
\end{eqnarray}

The same results was obtained previously\cite{VG},\cite{KP},\cite{LW}
and in particular  by  M.B.Green, P.Vanhove,
M.Gutperle and others in refs\cite{VG} calculating the one loop 
graviton scattering 
in the eleven dimensional supergravity compactified on $T^2$ after 
a regularization of the  UV divergent term proportional to  $\iT $
and assuming  
the validity of  Type IIA $\leftrightarrow$ Type IIB string  duality.

In our case the calculations are done without any assumption at all.
The result is UV-finite giving rise to an $SL(2,Z)$ invariant answer
as expected from the type IIB string theory and Type IIA 
$\leftrightarrow$ Type IIB  duality.

\boldmath
\section{Conclusions}
\unboldmath
We presented a suitable  conformal system of a  String and  a  Particle 
in $D=(10+2)$ dimensions. We show by using the  extra symmetries of this
system that one of the two time-like coordinates can be eliminated from
all the  remaining string observables and leads to  an effective string 
dimensionality
$D_{string}=(10+1)$. The 11th dimension arises after the resumming 
all the  topological non-trivial $(p,q)$ string sectors. These sectors 
appear as solutions of the orthogonality constrain between string and 
particle compactified momenta.

It is interesting that all perturbative and non-perturbative BPS
states in nine dimensions appear with masses parametrized
by the geometrical toroidal  moduli of  $T^2$.
In the $R^4$ calculations only the perturbative and non-
perturbative  BPS states contributed giving the  finite result
conjectured in refs \cite{VG},\cite{KP},\cite{LW} terms of the  Eisenstein $E_{3/2}(U)$
function:
$$
 \left(~ \iT  {\Large \pi \over 3 } ~+~
\iT^{-1/2}~{\zeta(3)
\over 2\pi}~E_{3/2}(U)~ \right)~ R^4
$$

In the  String $\&$ Particle Theory all 
the $R^{n}$  terms with  $ n > 4$
are determined unambiguously,  as in any conventional string theory,
by replacing the $\Gamma_{1,1}$ Lattice  of strings with the 
$\Gamma_{2,2}(p,q)$ restricted lattice of the  String $\&$ Particle
theory. In the $R^{n}$  with  $ n > 4$ terms, non BPS-states
(perturbative and non-perturbative)  give non-zero contribution
(contrary to $R^4$-terms). The String $\&$ Particle theory gives 
the correct masses for all these states and the multiplicities
are provided by the string oscillators. 

\vskip 1.cm
\centerline{\bf Acknowledgements}
\noindent
We thank P. M. Petropoulos, B. Pioline and P. Vanhove
for valuable discussions.

This work was partially supported by the EEC under the contract
TMR-ERBFMRX-CT96-0045.

\end{document}